\documentstyle[prl,aps,epsf,floats]{revtex}

\begin{document}
\draft
\twocolumn[\hsize\textwidth\columnwidth\hsize\csname @twocolumnfalse\endcsname
\title{
Current rectification by simple molecular quantum dots:  an ab-initio
study 
}

\author{B. Larade and A.M. Bratkovsky 
}
 
\address{
Hewlett-Packard Laboratories, 1501 Page Mill Road, Palo Alto, 
California 94304
}

\date{April 11, 2003}
\maketitle

\begin{abstract}

We calculate a current rectification by molecules containing
a conjugated molecular group sandwiched between two saturated (insulating)
molecular groups of different length (molecular quantum dot) using an ab-initio
non-equilibrium Green's function method.
In particular, we study S(CH$_2$)$_m$-C$_{10}$H$_6$-(CH$_2$)$_n$S
dithiol with Naphthalene as a conjugated central group.
The rectification current ratio $\sim 35$ has been observed at $m = 2$ and $n = 10$,
due to resonant tunneling through the molecular orbital (MO) closest to 
the electrode Fermi level
(lowest unoccupied MO in the present case). The rectification is limited by  
interference of other conducting orbitals, but can be improved by
e.g. adding an electron withdrawing group to the naphthalene.

\end{abstract}

\narrowtext
\pacs{PACS numbers: 85.65.+h}
\vskip2pc]

\section{Introduction}

Effective current rectifier is a necessary element of electronics circuitry.
Aviram and Ratner suggested in 1974 that in a molecule containing donor ($D)$
and acceptor ($A)$ groups separated by a saturated $\sigma $-bridge
(insulator) group, the (inelastic) electron transfer will be more favorable
from $A$ to $D,$ rather than in the opposite direction\cite{Aviram}. It was
noted more recently that the electron transfer from $D$ to $A$ should start
at smaller bias voltage because the highest occupied molecular orbital
(HOMO) is centered on $D$ group, whereas the lowest unoccupied molecular
orbital (LUMO) is located on the acceptor group $A$. The onset of resonant
tunneling in this case corresponds to the alignment of HOMO on the $D$ group
with LUMO on the $A$ group under external bias voltage, which are closest in
energy\cite{Ellenbogen}. The molecular rectifiers (MR) synthesized in the
laboratory were of somewhat different $D-\pi -A$ type, i.e. the ``bridge''
group was conjugated \cite{Martin,Metzger}. The molecule, C$_{16}$H$%
_{33}-\gamma $Q3CNQ, can be viewed as the fused naphthalene and TCNQ
molecules with attached C$_{16}$H$_{32}$ alkane ``tail'' needed to provide
enough van-der-Waals attraction between the molecules to assemble them into
Langmuir-Blodgett film on a water. Although the molecule did show a
rectification (with considerable hysteresis), it performed not like a
conductor but rather like an anisotropic insulator because of that large
alkane ``tail''. Indeed, the currents reported in \cite{Metzger} were on the
order of $10^{-17}$ A/molecule, which make its application 
impractical.
It was recently realized that in this molecule the
resonance does not come from the alignment of HOMO and LUMO, since they
cannot be decoupled through the conjugated $\pi -$bridge, but rather due to
an asymmetric voltage drop across the molecule where HOMO and LUMO are
asymmetrically positioned with respect to the Fermi level of the electrodes 
\cite{Krzeminski}. It is not clear if this mechanism has been observed. For
instance, the contact between the alkane chain and electrode was certainly
poor, via weak van-der-Waals forces. No temperature dependence of the
resistivity was reported, but its large value may suggest that inelastic
tunneling processes might have been involved.

It is clear that the rectifying molecular films built with the use of LB\
technique relying on use of long aliphatic (or other insulating) ``tails''
will produce hugely resistive molecules, not suitable for moletronics
applications. There are reports of rectifying behavior in other classes of
molecules, e.g. molecules chemisorbed on electrodes, where an observed
asymmetry of I-V curves is likely due to asymmetric contacts with electrodes 
\cite{Zhou,Xue}, or asymmetry of the molecule itself \cite{reichert}.

To achieve a good rectification by a molecule with a reasonable
conductivity, one should avoid using molecules with long saturated
(insulating) groups, which would make them prohibitively resistive. Thus, a
suggestion to use relatively short molecules with certain end groups to
allow their self-assembly on a metallic electrode's surface seems to be very
attractive\cite{KBWmr02}. In that previous paper the transport through a
phenyl ring connected to electrodes by asymmetric alkane chains (CH$%
_{2})_{n} $ has been calculated with the use of a semi-empirical
tight-binding method. Since the microscopic parameters of the model are
poorly known, especially hopping integrals between molecule and electrodes,
electrode work function and the affinity of the molecule (which may be
strongly affected by bonding to a metal), the calculations have been
performed for a number of these parameters. The calculations indicated a
rectification ratio for -S-(CH$_{2})_{2}$-C$_{6}$H$_{4}-$(CH$_{2})_{n}-$S-
molecule of about 100 for $n=10$ with the resistance $R\approx 13$G$\Omega .$
It is difficult to know how quantitative these numbers are, because of a
limited basis set and uncertainty in parameters of the model. For instance,
an account for larger basis set for wave functions may lead to an increased
estimate of the tunnel current, since the system would have more channels to
conduct current. Experimentally, it may be easier to synthesize somewhat
different compounds with a naphthalene group as a ``molecular quantum dot'',
like $-$S-(CH$_{2})_{m}$-C$_{10}$H$_{6}-$(CH$_{2})_{n}-$S$-$\cite{shunchi},
which we will also refer to as $-$S-(CH$_{2})_{m}$-Naph$-$(CH$_{2})_{n}-$S$%
-. $ To obtain an accurate description of transport in this case, we employ
an ab-initio non-equilibrium Green's function method\cite{Guo01}.

The calculation shows that indeed the current rectification $I_{+}/I_{-}\sim
100$ may be possible for some designs like $-$S-(CH$_{2})_{3}$-Naph$-$(CH$%
_{2})_{10}-$S$-$, where $I_{+(-)}$ is the forward (reverse) current. The
difference between the forward and reverse voltages is, however, limited by
other orbitals intervening into the conduction process. One needs the
conducting orbital to be much closer in energy to the electrode Fermi
level than the 
other ones (e.g. LUMO\ versus HOMO) and this energy asymmetry can be
manipulated by ``doping'' the conjugated conducting part by attaching a
donor (or acceptor)\ group, as will be shown below.

The present calculation takes into account only elastic tunneling
processes. Inelastic processes may substantially modify results in the
case of strong interaction of electrons with molecular vibrations,
see \cite{AB03vibr}. There are indications in the literature that the
carrier might be trapped in a polaron state in saturated molecules slightly
longer than we consider in the present paper\cite{Boulas}.

\section{Rectification by molecular quantum dot}

The structure of the present molecular rectifier is shown in
Fig.~1. The molecule consists of a central conjugated part
(naphthalene) isolated from the electrodes by two insulating ``arms''
built from saturated aliphatic chains (CH$_2$)$_n$ with lengths $L_1$
($L_2$) for the left (right) chain.
The principle of molecular rectification by a molecular quantum dot is
illustrated in Fig.~2, where the electrically ``active'' molecular orbital,
localized on the middle conjugated part, is LUMO, which lies at the energy $%
\Delta $ above the electrode Fermi level at zero bias. The position of the
LUMO, for instance, is determined by the work function of the metal $q\phi $
and the affinity of the molecule $q\chi,$ $\Delta =\Delta _{LUMO}=q\left(
\phi -\chi \right) .$ The position of HOMO\ is given by $\Delta
_{HOMO}=\Delta _{LUMO}-E_{g},$ where $E_{g}$ is the HOMO-LUMO\ gap. If this
orbital is considerably closer to the electrode Fermi level $E_{F},$ then it
will be brought into resonance with $E_{F}$ prior to other orbitals when the
molecule is biased in either forward or reverse direction (Fig.~2). It is
easy to estimate the forward and reverse bias voltage assuming that the
voltage mainly falls off at the saturated (wide band gap) parts of the
molecule with the lengths $L_{1}$ ($L_{2})$ for the left (right) barrier,
Fig.~2. Since the voltage drops proportionally to the lengths of the
barriers, $V_{1(2)}\propto L_{1(2)},$ we obtain for the partial voltage
drops $V_{1(2)}=VL_{1(2)}/L$, where $V$ is the bias voltage. Forward and
reverse voltages are found from the resonance condition, which gives 
\begin{figure}[t]
\epsfxsize=3.4in \epsffile{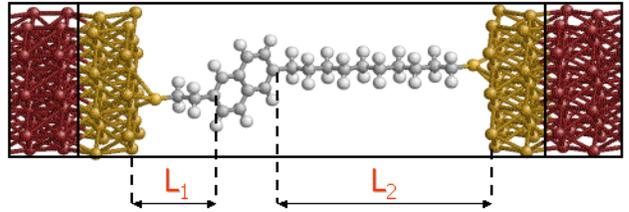}
\caption{Schematic diagram of typical Au-molecular rectifier-Au device
structure analyzed within our ab-initio formalism. Electrodes are
constructed as the 111 orientation, with the terminal sulfurs of the
molecule occupying the triangular hollow site. $L_{1}$ ($L_{2})$ indicates
the insulating barrier length separating the molecular conducting unit from
the left (right) electrode.
}
\label{fig:molec}
\end{figure}
\begin{figure}[t]
\epsfxsize=2.5in \epsffile{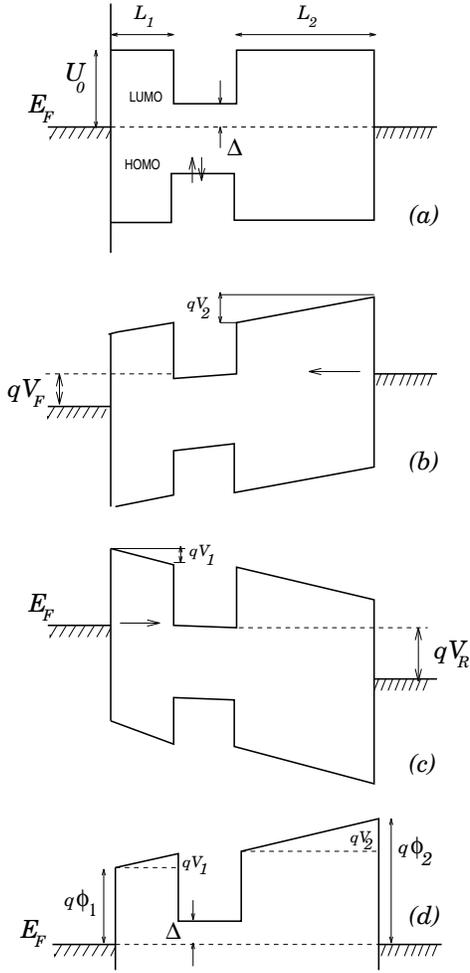}
\caption{Schematic band diagram of the molecular rectifier. Middle narrow-band
part represents the naphthalene conjugated group. The barrier height
$U_0 = 4.9$ eV (present theory), $4.8$eV (experiment [16]).
(a)-(c) Unbiased, forward, and reverse biased rectifier with similar
electrodes. (d) Unbiased rectifier with dissimilar electrodes. At
equilibrium, there is a voltage drop $qV_{1(2)}$ across the left
(right) barrier. 
}
\label{fig:bands}
\end{figure}
\begin{eqnarray}
V_{F} &=&\frac{\Delta }{q}(1+\xi ), \\
V_{R} &=&\frac{\Delta }{q}\left( 1+\frac{1}{\xi }\right) , \\
V_{F}/V_{R} &=&\xi \equiv L_{1}/L_{2},  \label{eq:vratio}
\end{eqnarray}
where $q$ is the elementary charge. A significant difference between forward
and reverse currents should be observed in the voltage range $%
V_{F}<|V|<V_{R} $. For our systems in question the parameter $\Delta
=1.2-1.5 $ eV, Fig.~3, 
\begin{figure}[t]
\epsfxsize=3.2in \epsffile{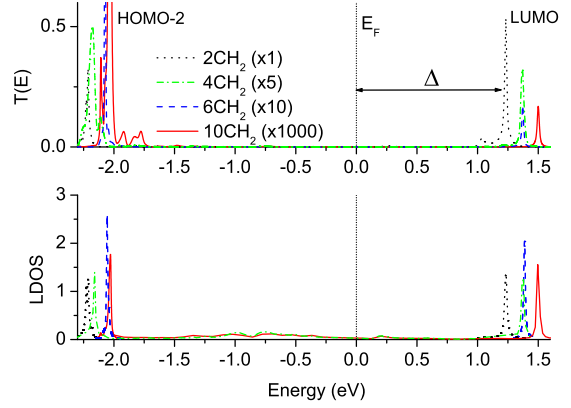}
\caption{
Transmission coefficient (top) and local density of states
(bottom) versus energy E for rectifiers $-$S-(CH$_{2})_{2}$-C$_{10}$H$_{6}-$%
(CH$_{2})_{n}-$S$-,$ $n=2,4,6,10$. $\Delta $ indicates the distance of the
closest MO to the electrode Fermi energy (E$_{F}$ = 0).
}
\label{fig:fig3}
\end{figure}
\begin{figure}[t]
\epsfxsize=3.2in \epsffile{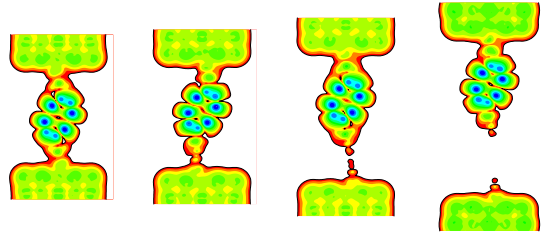}
\caption{
 Projected density of states $N(E,\vec r)$, onto a 2-D grid defined by the plane
of our central molecule, for the naphthalene series, $n=2,4,6,10$ from left
to right, respectively. As the alkane chain increases in length, the
coupling of the naphthalene group to the right electrode decreases.
}
\label{fig:fig4n}
\end{figure}
the value determined by the work function of the
electrodes and electron affinity of the conjugated part of the molecule, as
well as by the strength of a surface dipole layer. For unlike electrodes
with the work functions $q\phi _{1}$ and $q\phi _{2}$ there will be a
built-in potential in the system $q\phi _{b}=q\left( \phi _{2}-\phi
_{1}\right)$, Fig.~2(d). In this case the expressions for the forward
and reverse bias 
voltage are modified as follows$:$%
\begin{eqnarray}
V_{F} &=&\left( \frac{\Delta }{q}+\phi _{b}\right) (1+\xi ), \\
V_{R} &=&\phi _{b}+\frac{\Delta }{q}\left( 1+\frac{1}{\xi }\right) .
\end{eqnarray}
Thus, the operating bias ``window'' in this case 
\begin{eqnarray}
V_{R}-V_{F} &=&\left( V_{R}-V_{F}\right) _{0}-\xi \phi _{b}  \nonumber \\
&=&\frac{\Delta }{q}\left( \frac{1}{\xi }-\xi \right) -\xi \phi _{b}.
\label{eq:wind}
\end{eqnarray}
Therefore, it slightly ($\xi <1$) {\em increases} when $\phi _{b}<0$ ($\phi
_{1}>\phi _{2}$, left work function is larger$),$ and {\em shrinks down }%
when $\phi _{b}>0$ (right work function is larger). One would like to use as
asymmetric a molecule as possible to increase the operating voltage, and in
this case the difference of the work functions of the electrodes
progressively becomes less important.

The main transport features of the present system can be qualitatively
explained within a model of resonant tunneling through a localized state
(molecular orbital) in the barrier. The current can be written as 
\begin{equation}
I=\frac{2q}{h}\int dET(E)\left[ f(E)-f(E+qV)\right] ,
\label{eq:cur}
\end{equation}
where the transmission $T(E)\;$at zero temperature is given by (for two spin
directions) 
\begin{equation}
T(E)=\frac{\Gamma _{L}\Gamma _{R}}{\left( E-E_{MO}\right) ^{2}+\left( \Gamma
_{L}+\Gamma _{R}\right) ^{2}/4},  \label{eq:TEbw}
\end{equation}
where $E_{MO}=E_{F}+\Delta -qV_{1}=E_{F}+\Delta -qVL_{1}/L$ is the energy of
the molecular orbital, relative to the Fermi level of the left electrode.
The transmission is maximal and equals $1$ when $E=E_{MO}$ and $\Gamma
_{L}=\Gamma _{R},$ which corresponds to a symmetric position of the central
conjugated part with respect to the electrodes. We can estimate 
$\Gamma_{L(R)}\sim t^{2}/D=\Gamma _{0}e^{-2\kappa L_{1(2)}},$ 
where $t$ is the
overlap integral between the MO\ and the electrode, $D$ is the electron band
width in the electrodes, $\kappa $ the inverse decay length of the
resonant MO into the barrier. The latter quantity can be estimated from $%
\hbar ^{2}\kappa ^{2}/2m^{\ast }=U_{0},$ where $U_{0}$ is the barrier
height [$\approx 4.8$ eV in alkane chains (CH$_{2})_{n}$\cite{Boulas}, {\bf %
\ }$\approx 5$ eV from DFT calculations], and $m^{\ast }\lesssim 1$ the
effective tunneling mass. The current above the threshold (when the resonant
tunneling condition is met)\ is given by 
\begin{equation}
I\approx \frac{2q}{\hbar }\Gamma _{0}e^{-2\kappa L_{2}}.
\end{equation}
We see immediately that increasing the spatial asymmetry of the design (i.e.
increasing $L_{2}/L_{1})$ changes the operating voltage range linearly, $%
V_{R}-V_{F}\approx \left( \Delta /q\right) L_{2}/L_{1},$ and brings about an 
{\em exponential} decrease in current\cite{KBWmr02}. This severely limits
the ability to increase the rectification ratio while simultaneously keeping
the resistance of the molecule at a reasonable value. One therefore needs a
quantitative analysis of all microscopic details, including
molecule-electrode contact, and potential and charge distribution, in order
to make comparisons with the data.

\section{Method}

Since we study relatively short molecules, electron transport is likely to
proceed elastically. We shall treat the elastic tunneling processes from
first principles in order to evaluate all the relevant microscopic
characteristics within the density functional theory. This allows us to
check the accuracy of semi-empirical methods used previously, and find
results for new objects of experimental interest, like the present molecular
quantum dot with a naphthalene conjugated middle group. We use an ab-initio
approach that combines the Keldysh non-equilibrium Green's function (NEGF) 
\cite{jauho94} with pseudopotential-based real space density functional
theory (DFT)\cite{Guo01}. In contrast to standard DFT work, our NEGF-DFT
approach considers a quantum mechanical system with an {\it open }boundary
condition provided by semi-infinite electrodes under external bias voltage.
\ The main advantages of our approach are (i) a proper treatment of the open
boundary condition; (ii) a fully atomistic treatment of the electrodes and
(iii) a self-consistent calculation of the non-equilibrium charge density
using NEGF.

For the open devices studied here, our system is divided into three regions,
a left and right electrode and a central scattering region, which contains a
portion of our semi-infinite electrodes (see Fig.~1). \ By including atomic
layers of the electrode in our scattering region, we provide a proper
treatment of the electrode-molecule interaction, and allow for the effective
screening of the molecule away from our molecular junction. \ External bias
voltage can then be applied as a boundary condition for the Poisson equation
in our central scattering region. \ The Hartree potential in the scattering
region is solved using a multigrid real-space numerical approach. \ Finally,
the non-equilibrium density matrix $\hat{\rho}$ is constructed via NEGF \cite
{jauho94}, 
\begin{equation}
\hat{\rho}=\frac{1}{2\pi i}\int dE{\rm Tr}G^{<}(E),
\end{equation}
where the lesser Green's function is determined from the operator relation 
\begin{equation}
G^{<}=G^{R}\Sigma ^{<}G^{A},
\end{equation}
via the retarded (advanced) Green's functions $G^{R(A)}$ and $\Sigma ^{<}$
the self-energy part describing an injection of charge into the scattering
region (molecule)\ from the electrodes. This process is characterized by the
usual {\em open channel} representation of the scattering states with
momenta $k_{l}^{n}$ ($k_{r}^{n})$ in the left (right)\ electrodes, where
index $n$ enumerates all open channels (available Bloch states of the
electrons in the electrodes). The Fermi functions $f(k_{l}^{n})$ and $%
f(k_{r}^{n})$ define the occupied states in the leads to be accounted for in
evaluation of the electron transport. To evaluate the Green's function, one
treats the {\em disconnected} system consisting of the left (right) lead
marked by index $l$ $(r)$ and the scatterer $c.$ The transport Green's
function is then found from the Dyson equation 
\begin{equation}
\left( G^{R}\right) ^{-1}=\left( G_{0}^{R}\right) ^{-1}-V,
\end{equation}
where the unperturbed retarded Green's function is defined in operator form
as 
\begin{eqnarray}
\left( G_{0}^{R}\right) ^{-1} &=&(E+i0)\hat{S}-\hat{H}, \\
\hat{H} &=&\left( 
\begin{array}{ccc}
H_{l,l} & H_{l,c} & 0 \\ 
H_{c,l}  & H_{c,c} & H_{c,r} \\ 
0 & H_{r,c} & H_{r,r}
\end{array}
\right) .
\end{eqnarray}
$H$ is the Hamiltonian matrix, $H_{c,l}=H_{l,c}^\dagger $,
$H_{r,c}=H_{c,r}^\dagger $, 
 $S$ is the {\em overlap} matrix, $%
S_{i,j}=\left\langle \chi _{i}|\chi _{j}\right\rangle $ for non-orthogonal
basis set orbitals $\chi _{i}$, and the coupling of the scatterer to the
leads is given by the Hamiltonian matrix $V={\rm diag}[\Sigma
_{l,l},~0,~\Sigma _{r,r}].$ In all formulas above it is assumed that
matrix elements are given in terms of the non-orthogonal basis set. 

The self-energy part $\Sigma ^{<},$ which is used to construct the
non-equilibrium electron density in the scattering region, is
found from 
\begin{equation}
\Sigma ^{<}=-2i%
\mathop{\rm Im}%
\left[ f(E)\Sigma _{l,l}+f(E+qV)\Sigma _{r,r}\right] .
\end{equation}
This expression accounts for the steady charge ``flowing in'' from the
electrodes. The transmission probability, which determines the current
according to Eq.~(\ref{eq:cur}), is given by 
\begin{equation}
T(E)=4{\rm Tr}\left[ 
\mathop{\rm Im}%
\left( \Sigma _{l,l}\right) G_{l,r}^{R}%
\mathop{\rm Im}%
\left( \Sigma _{r,r}\right) G_{r,l}^{A}\right].
\end{equation}
The local density of states, which is a very important quantity to
characterize the spatial character of the wave functions, can be obtained
from 
\begin{equation}
N(E,\vec{r})=\frac{1}{2\pi i}{\rm Tr}\left[ G^{<}(E)\left| \chi (\vec{r}%
)\right\rangle \left\langle \chi (\vec{r})\right| \right] .
\end{equation}
To construct the system Hamiltonian, we define the atomic cores using
standard norm-conserving nonlocal pseudopotentials\cite{trou91}, and expand
the Kohn-Sham wave functions in an s-,p-,d- real space atomic orbital basis 
\cite{Guo01,ordejon96}. External bias is incorporated into the Hartree
potential, and in this way the non-equilibrium charge density is iterated to
self-consistency, after which the current-voltage (I-V) characteristics have
been computed.

\section{Results}

The physical picture of the transmission through a series of rectifiers $-$%
S-(CH$_{2})_{m}$-C$_{10}$H$_{6}-$(CH$_{2})_{n}-$S$-$ for $m=2$ and $%
n=2,4,6,10$ is illustrated in Figs.~3,4. The transmission
coefficient $T(E)$ (top panel) and the density of states $N(E)=\int
dVN(E,\vec{r})$ (bottom panel) are shown at 
zero bias voltage in Fig.~3. The Fermi level in the electrodes is
taken as the energy 
origin. It follows from this figure that indeed the LUMO\ is the closest
molecular orbital transparent to electron transport, and it is higher
in energy than $E_{F}$ by an amount $\Delta =1.2-1.5$ eV depending on the
molecule. Since the molecule is asymmetric, even at the resonance energy we
have $T(E_{LUMO})\lesssim 0.5,$ Fig.~3. 
\begin{figure}[t]
\epsfxsize=3.in \epsffile{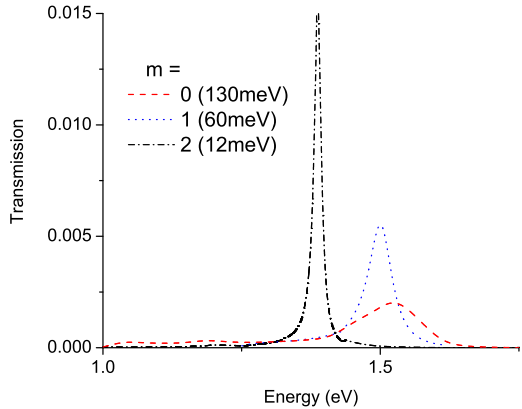}
\caption{Transmission through molecular rectifiers $-$S-(CH$_{2})_{m}$-C$%
_{10} $H$_{6}-$(CH$_{2})_{6}-$S$-,$ $m=0,1,2$. \ The peak corresponds to
transmission through the LUMO\ state localized on the naphthalene ring.
}
\label{fig:fig5n}
\end{figure}
This can be traced back to the
Breit-Wigner formula (\ref{eq:TEbw}), which for $\Gamma _{R}<\Gamma _{L}$
gives $T(E_{LUMO})\approx 4\Gamma _{R}/\Gamma _{L}<1,$ and it falls off as
the asymmetry increases. There the transmission
through nominal HOMO\ state is negligible, but HOMO-2 state conducts very
well. As we shall see below, the HOMO-2 defines the threshold reverse
voltage $V_{R},$ thus limiting the operating voltage range
(\ref{eq:wind}).

The maps of the densities of states $N(E_{LUMO},\vec{r})$ in Fig.~4 show
the development of the tunneling gap between the naphthalene group and more
distant electrode separated by alkane chain (CH$_{2}$)$_{n},$ $n=2-10.$ The
symmetry and weight distribution of LUMO in the device is very similar
to LUMO in the isolated molecule, telling us that the character of the
molecular wave functions is preserved to a high degree. Obviously, in
contact with electrodes molecular orbitals acquire some finite width $\Gamma
=\Gamma _{L}+\Gamma _{R},$ and for highly asymmetric molecules $\Gamma
\approx \Gamma _{L}$, so it is mainly defined by the transparency of the
shorter (left)\ insulating group (CH$_{2}$)$_{m}$. One needs a rather small
width $\Gamma $ which defines the sharpness of the voltage threshold for the
current turn-on \cite{KBWmr02}. We have calculated the molecules with $%
m=0,1,2$ with the results for $\Gamma _{m}=130,$ $60$, and 12 meV (Fig.~5).
The last value for the left insulating group (CH$_{2})_{2}$ seems to be most
reasonable, since the resonant peak is rather narrow yet the molecule
remains conductive. However, in some cases the short insulating group (CH$%
_{2})_{3}$ may be preferential. This conclusion is similar to the one
reached on the basis of the tight-binding model, but the peak positions and
size are substantially different in both cases, the peak size being much
larger in the case of the present DFT\ calculations, Fig.~5.

Our assumption that the voltage drop is proportional to the lengths of the
alkane groups on both sides, is quantified by the calculated potential ramp
in Fig.~6. 
\begin{figure}[t]
\epsfxsize=3.2in \epsffile{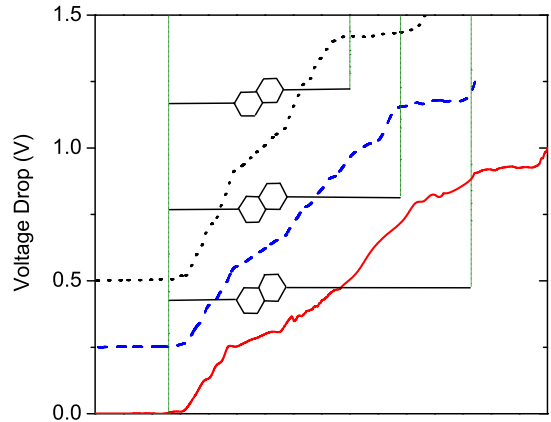}
\caption{Potential drop as a function of position along the length (transport
direction) of our molecular device, for the $n=2,6,10$ molecular rectifiers,
at an applied bias voltage of 1V. The positions of the naphthalene ring
inside the junction are indicated by the stick figures, and the respective
insulating barrier lengths are shown as horizontal lines. Curves are
vertically shifted for clarity.
}
\label{fig:fig6n}
\end{figure}
It is close to a linear slope along the (CH$_{2})_{n}$ chains.
There is also a smaller but still noticeable voltage drop across the
naphthalene group, which can be viewed as a fragment of wide-band
semiconductor.

The threshold bias voltages can be obtained from Fig.~7,
\begin{figure}[t]
\epsfxsize=3.2in \epsffile{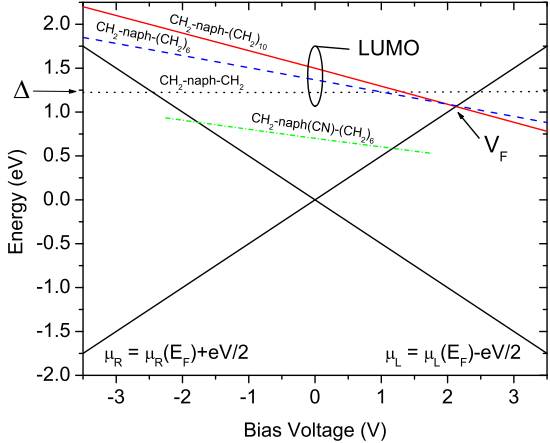}
\caption{
Evolution of the rectifier LUMO energy state as a function of
applied bias voltage. Solid black lines represent the electrode chemical
potentials, which are symmetrically shifted by the applied voltage $\mu
_{R(L)}=E_{F}+(-)qV/2$. The forward voltage threshold corresponds to the
crossing of the LUMO energy with the right electrode chemical potential.
In the case of CN-doped naphthalene ring both forward and reverse
threshold voltages correspond to LUMO crossing the electrode Fermi
level, as in the ideal case, Fig.~2. In other systems the HOMO defines
the reverse 
voltage threshold, thus reducing the operating voltage range.
}
\label{fig:fig7n}
\end{figure}
 showing the
evolution of the LUMO\ versus bias voltage with respect to the chemical
potentials of the right (left) electrodes $\mu _{R(L)}=E_{F}+(-)qV/2.$ The
forward voltage corresponds to the crossing of the LUMO($V)$\ and $\mu
_{R}(V)$, which happens at about 2V. Extrapolated crossing of LUMO\ and $\mu
_{L}$ is at large negative voltages but, unfortunately, this large desirable
difference between $V_{F}$ and $V_{R}$ does not materialize. Although the
LUMO\ defines the forward threshold voltages in all the molecules studied
here, the reverse voltage is defined by the HOMO-2\ for ``right'' barriers
(CH$_{2})_{n}$ with $n=6,10$. The I-V curves are plotted in Fig.~8 
\begin{figure}[t]
\epsfxsize=3.2in \epsffile{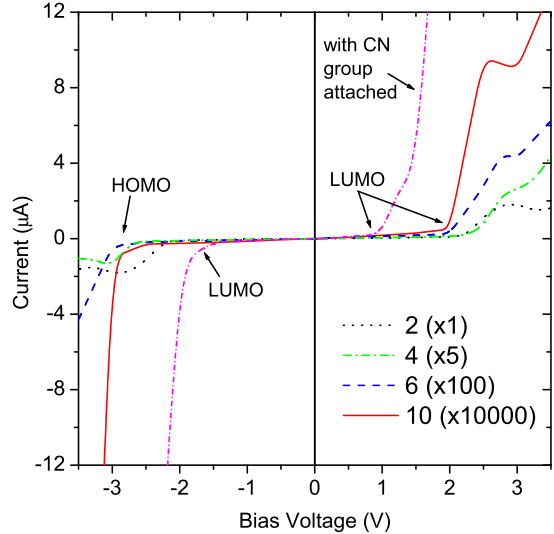}
\caption{I-V curves for naphthalene rectifiers $-$S-(CH$_{2})_{2}$-C$_{10}$H%
$_{6}-$(CH$_{2})_{n}-$S$-,$ $n=2,4,6,10$. \ The short-dash-dot curve
corresponds to a cyano-doped (added group -C$\equiv $N) $n=6$ rectifier.
}
\label{fig:fig8n}
\end{figure}
and the
corresponding parameters are listed in the Table. We see that the
rectification ratio for current in the operation window $I_{+}/I_{-}$
reaches the maximum value of 35 for ``2-10'' molecule ($m=2,$ $n=10).$ It
was estimated to be about 100 for ``2-10'' molecule with a phenyl ring as a
conjugated central part\cite{KBWmr02}. We have checked a series of molecules
with central phenyl ring and, unfortunately, do not find any significant
rectification in this case, because the Fermi level in DFT\ calculations
appears to lie close to mid-gap. Thus, the tight-binding results for large
rectification\ in one-phenyl ring molecules might have been the results of
the model. The estimated resistance of the molecule in DFT\ is substantially
less (2.8 G$\Omega )$ compared to the tight-binding model with a phenyl ring
as a conjugated central unit (13 G$\Omega )$, see Table.
\begin{table}
\caption{The parameters of a set of the naphthalene-based rectifiers
with various right insulating groups (CH$_2$)$_n$, Fig.~1. $I_{+}/I_{-}$ the
currect rectification ratio, R the molecule resistance at bias voltage
$V=2.5$V.}
\begin{tabular}{lccccc}
$n$ & $\Delta$ (eV) &
$V_F$ (V) & $V_R$ (V) & $I_{+}/I_{-}$ & $R$ (M$\Omega$)\\ \tableline\tableline
2 & 1.23 & 2.45 & 2.45 & 1 & 2.20 \\
4 & 1.37 & 2.50 & 3.00 & 5 & 20.1 \\
6 & 1.39 & 2.15 & 3.13 & 15 & 90.9 \\
10 & 1.50 & 2.10 & 2.85 & 35 & 2780 \\
\end{tabular}
\end{table}

One can manipulate the system in order to increase the energy asymmetry of
conducting orbitals (reduce $\Delta )$. To shift the LUMO\ towards $E_{F}$,
one needs in effect to gate the naphthalene group by positive voltage. A
chemical way of doing this would be to attach some chemical group, like -C$%
\equiv $N in the present case. This group withdraws a fraction of an
electron and acquires the negative charge $-\delta =$ $-0.11q.${\bf \ }The
charge is donated mainly by the naphthalene group, which acquires a positive
charge and orbitals on the molecule shift to lower energies almost rigidly
by about 0.6 eV, Fig.~9. 
\begin{figure}[t]
\epsfxsize=3.2in \epsffile{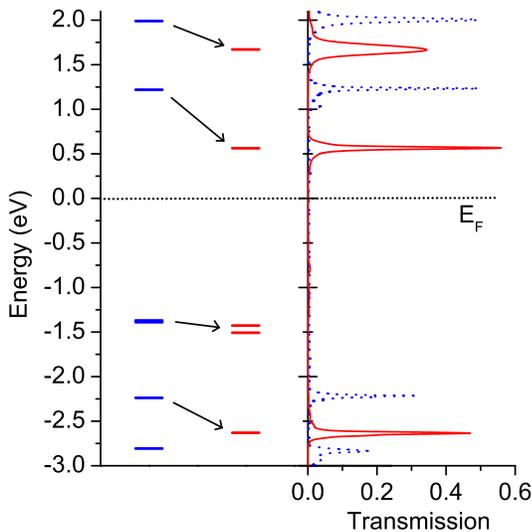}
\caption{
 Left: \ Energy level positions for undoped and cyano doped (-C$%
\equiv $N) symmetric (m,n = 2) naphthalene rectifier. \ Right: \
Transmission $T(E)$ for the doped (solid) and undoped (dotted) rectifiers at
zero applied bias voltage. \ The -C$\equiv $N substituent shifts the
electronic levels near $E_{F}$ to lower energy, increasing the energy
asymmetry and hence decreasing $\Delta .$
}
\label{fig:fig9n}
\end{figure}
On the resulting I-V curve, both thresholds $%
V_{F(R)}$ now correspond to LUMO\ crossing either the right or left
electrode Fermi levels (see Fig.~7), and the operating voltage range is
modified significantly.

\section{Conclusions}

We have presented parameter-free DFT\ calculations of a class of molecular
quantum dots showing a current rectification at a reasonable current density
through the molecule. The explored mechanism of rectification is different
from Aviram-Ratner diodes\cite{Aviram} and relies on considerable asymmetry
of the spatial composition and energy structure of conducting orbitals of
the molecule\cite{KBWmr02}. Since there is no empirical parameters, the
predictions are quantitative and can be tested in a controllable fashion\cite
{shunchi}. We have studied the molecular quantum dots with the naphthalene
central conjugated group, and found a rectification of about 35 for $-$S-(CH$%
_{2})_{2}$-C$_{10}$H$_{6}-$(CH$_{2})_{6}-$S$-$ molecules. The modest
rectification ratio is a consequence of the resonant processes at these
molecular sizes. The present study suggests also that the rectification
ratio in the case of a central single phenyl ring, studied previously\cite
{KBWmr02}, is dramatically overestimated by the tight-binding model{\bf .}

The rectification ratio is not great by any means, but one should bear in
mind that this is a device necessarily operating in a ballistic
quantum-mechanical regime, because of small size. This is very different
from present Si devices with carriers diffusing through the system. As they
become smaller, however, the same effects, as those discussed here for
molecular rectifiers, will eventually take over and will tend to diminish
the rectification ratio. In this regard the present results are indicative
of the problems which will ultimately be faced by Si devices of molecular
proportions.

It is important that the parameters of the device, like the threshold
voltages for forward and reverse bias, can be significantly modified by
``chemical'' doping. In the present case adding the -C$\equiv $N electron
withdrawing group has scaled the threshold range by a factor of about two.
There may be other modifications that could improve/add functionality of the
molecular devices by varying central, end, and side groups of the molecule.
The present results illustrate very clearly that the optimization of the
parameters of the molecular devices is typically facing the problem that
some parameters, like level widths, vary exponentially, while others, like
the voltage thresholds, vary as a power law-- as a function of the molecular
design. Although the approximate band diagrams illustrating the level
structure, Fig.~2, are very helpful for qualitative description, the
atomistic details of the electron density distribution are extremely
important, Fig.~4. This necessitates the use of ab-initio methods,
without which one may obtain qualitatively wrong results.

We have found that in the present molecules significant rectification may be
obtained in relatively short self-assembled molecules. Self-assembly
provides for good contact with electrode and relatively small resistance per
molecule. In existing studies of Aviram-Ratner-type molecules they were
assembled by Langmuir-Blodgett deposition and are prohibitively resistive by
design because of a long aliphatic tail needed for the assembly \cite
{Martin,Metzger}. Thus, for applications the self-assembly of shorter
molecules on metallic electrodes seems to be essential. One should keep in
mind that the molecular rectification, here estimated for ideal contacts at
zero temperature, will be reduced at finite temperature and due to
inevitable disorder in the contact area and in the molecular film\cite{BK02}%
. Indeed, the optimal width of the level was found to be about 12 meV, which
is just a half of the room temperature, so the temperature (and static
disorder) will significantly broaden the level and reduce the rectification.
In terms of simplest cross-point memories one may use a large number of
molecules on a pitch. This may allow for use of longer, better rectifying
molecules. There, however, the inelastic processes will play a progressively
more significant role\cite{Boulas} and will further reduce the
rectification. Further studies of molecular rectifiers should be focused on
these outstanding issues in conjunction with experimental studies.

We acknowledge useful discussions with Shun-chi Chang and
R. S. Williams. The work has been supported by DARPA.

\end{document}